# Competing Heterogeneities in Vaccine Effectiveness Estimation

Ariel Nikas[1], Hasan Ahmed[2], Veronika I. Zarnitsyna[1*]

[1]Emory University School of Medicine, Department of Microbiology and Immunology, Atlanta, Georgia, United States of America

[2]Emory University, Department of Biology, Atlanta, Georgia, United States of America

**\* Correspondence:**
Corresponding Author
vizarni@emory.edu

## Abstract

Understanding waning of vaccine-induced protection is important for both immunology and public health. Population heterogeneities in underlying (pre-vaccination) susceptibility and vaccine response can cause measured vaccine effectiveness (mVE) to change over time even in the absence of pathogen evolution and any actual waning of immune responses. We use a multi-scale agent-based models parameterized using epidemiological and immunological data, to investigate the effect of these heterogeneities on mVE as measured by the hazard ratio. Based on our previous work, we consider waning of antibodies according to a power law and link it to protection in two ways: 1) motivated by correlates of risk data and 2) using a within-host model of stochastic viral extinction. The effect of the heterogeneities is given by concise and understandable formulas, one of which is essentially a generalization of Fisher's fundamental theorem of natural selection to include higher derivatives. Heterogeneity in underlying susceptibility accelerates apparent waning, whereas heterogeneity in vaccine response slows down apparent waning. Our models suggest that heterogeneity in underlying susceptibility is likely to dominate. However, heterogeneity in vaccine response offsets <10% to >100% (median of 29%) of this effect in our simulations. Our methodology and results may be helpful in understanding competing heterogeneities and waning of immunity and vaccine-induced protection. Our study suggests heterogeneity is more likely to 'bias' mVE downwards towards faster waning of immunity but a subtle bias in the opposite direction is also plausible.

## 1    Introduction

Providing accurate estimates of vaccine-induced protection is essential in guiding public health policy. However, many factors complicate our ability to estimate population level vaccine effectiveness (VE) such as prior immunity, underlying health risks, timing of vaccination, inconsistent hazards in different locations, and other confounders. Further adding to uncertainty is the common presence of observed waning which may reflect actual waning of immune responses, introduction of different strains, or may be an artifact coming from heterogeneity among individuals.

Many studies report fast, intraseasonal waning of vaccine-induced protection, particularly for viruses such as influenza and SARS-CoV-2 (1-3); however, various effects can bias this conclusion. Depletion of susceptible individuals (also called the frailty effect in biostatistics) can bias estimates

(4, 5). Heterogeneity in exposure risk, even if exactly the same in the vaccinated and unvaccinated groups, tends to bias the vaccine effectiveness estimates downwards potentially leading to spurious claims of waning (6, 7). If natural immunity is not taken into account, merely having a "leaky" vaccine (i.e. a vaccine that provides partial protection) can bias estimates downwards (5, 7). This complicates the estimation of actual waning of vaccine-induced protection which is expected to occur as many correlates of protection, e.g. neutralizing antibodies, have been shown to decrease over time (8-11).

In this paper, we focus on the hazard ratio as the measure of vaccine effectiveness, as the hazard ratio corresponds to relative risk at a particular moment in time. To determine the direction and magnitude of bias, we simulate an epidemic in a population under various frailty and vaccine protection distributions in the absence and presence of waning and evaluate the interplay between these factors. Commonly, hazard ratios are estimated with the Cox proportional hazards model and there are several standard extensions utilized in real world studies (12-19). While Cox proportional hazards models were not intended to be time-varying, several approaches exist in order to make it applicable for measuring waning vaccine effectiveness. We utilize time category-vaccine interactions (henceforth TVI) which, in contrast to the commonly used Cox method utilizing the scaled Schoenfeld residuals, should accurately measure the hazard ratios even in the presence of extreme (observed) waning (20).

If vaccine effectiveness is assessed via the hazard ratio and the outcome of interest is the first infection post-vaccination, heterogeneity in baseline (pre-vaccination) susceptibility causes measured vaccine effectiveness (mVE) to decline over time, whereas heterogeneity in response to vaccination causes mVE to increase over time. Hence any apparent change in mVE may reflect any combination of selection on these heterogeneities in addition to the biological processes of pathogen evolution and waning of immune responses. In this paper, we first illustrate the problem using standard statistical distributions for the heterogeneities. We then provide concise formulas that give the net effect of these heterogeneities on the hazard rates and hazard ratio. Next, using parameter values based on epidemiological and immunological data incorporating waning of antibodies, we use agent-based simulations to investigate the magnitude of these opposing effects. Our models suggest that the larger effect is from heterogeneity in baseline susceptibility but that heterogeneity in vaccine response may offset a substantial fraction of that effect. This exacerbation of observed waning may explain the sometimes negative VE reported in some studies (21, 22).

## 2    Methods

We consider an agent-based model of acute viral infection with a constant background force of infection where we introduce heterogeneity in individual infection risk, heterogeneity in vaccine-induced protection, or both. For most scenarios to be described, vaccine protection follows the "leaky" model, wherein each vaccinated individual experiences a certain percent reduction in risk. Additionally, we model 40% out of a cohort of 100,000 to be vaccinated, in line with the CDC estimate for influenza vaccine coverage (23). Since we model an acute viral infection, we assume sterilizing immunity upon infection for the remainder of the one-year time period considered. All simulations were run in *Julia* version 1.3.1 and statistical analysis was completed in *R* version 4.2.1.

For heterogeneity in underlying risk (risk in the absence of vaccination), we select a daily risk rate for both the vaccinated and unvaccinated groups from a single gamma distribution. For heterogeneity in vaccine protection, we select vaccinated individuals' protection from a variety of distributions including beta distributions, in contrast to leaky, homogeneous vaccination. To establish a comparison, we use the mean vaccine efficacy in the context of no epidemic, $VE_{NE}$, which thereby removes the effect of selection. We then calculate vaccine effectiveness using a time category-vaccine-interaction (TVI) as the independent variable of the Cox proportional hazards model in order to find a time-varying estimate of protection. The TVI method has been shown to behave accurately in circumstances where waning occurs rapidly (20).

# 3 Results

## 3.1 Susceptibility to Infection (Frailty) Distribution

When considering only heterogeneity in underlying frailty, the given gamma distributions (parameterized as Gamma($\alpha_\gamma$, $\beta_\gamma$) where $\alpha_\gamma/\beta_\gamma$ is the mean and $\alpha_\gamma^{-0.5}$ is the coefficient of variation (CoV)) can produce the appearance of waning vaccine effectiveness though as $\alpha_\gamma$ increases this effect lessens. This appearance of waning corresponds to what many statistical studies have posited would occur and termed the "frailty effect" or "frailty phenomena" (24-27); in epidemiological studies, this is sometimes alternatively called "survivor bias" or "depletion of susceptibles" effect. Some studies have also simulated this effect but have not compared the qualitative effect of different distributions (5, 6). In Figure 1 we show how different gamma distributions with the same mean can cause differing amounts of perceived waning (waning in mVE) when the true vaccine protection is in fact constant and leaky.

## 3.2 Vaccine Efficacy Distribution

In simulated studies, two modes of vaccine efficacy are often compared: "leaky" vaccination where protection is incomplete but reduces each individual's chance to become infected by a specified amount (e.g. 50%) or "all-or-nothing" vaccination where a fraction of individuals receive complete protection from the vaccine and other receive no protection. A limited number of studies have also considered normal-like distributions for vaccine protection (7). We consider two main beta distributions parameterized by Beta($\alpha_\beta$, $\beta_\beta$) where $\alpha_\beta/(\alpha_\beta + \beta_\beta)$ is the mean (held here at 0.5): the normal-like Beta(2,2) and the U-shaped Beta(0.5,0.5). These distributions as well as their resultant dynamics and mVE estimates are compared in Figure 2 which shows that for both of these cases, distributions in vaccine protection bias VE estimates upwards. Non-symmetric vaccine protection distributions were also tested (see Supplement Figure S1) but did not change the direction of bias, showing increase in mVE.

As seen in Figures 1 and 2, singly the two types of heterogeneity appear to contribute in opposite directions; beta distributed vaccine protection tends to skew the estimate upwards while gamma distributed underlying risk tends to skew the estimates downwards and to a greater extent. As these effects compete when combined, we constructed a predictor to determine which direction, if any, the competing distributions change mVE from $VE_{NE}$.

## 3.3 Effect of Selection on Heterogeneities

Assuming hazard rates for a given individual are not time-varying and ignoring stochasticity, $\bar{r}$, the average hazard rate in not-yet-infected individuals, is given by the following equation

$$\bar{r}(t) = \frac{\int f(r) e^{-rt} r \, dr}{\int f(r) e^{-rt} \, dr}. \tag{1}$$

Here $f(r)$ is the probability density function for the hazard rates at time 0; we allow $f(r)$ to be a generalized function (e.g. a delta function) so the formula also applies to discrete probability distributions. Let $R$ denote the random variable for $f(r)$. Let $M(t)$ and $K(t)$ be the moment generating and cumulant generating functions of $-R$, respectively. Since the denominator of Equation 1 is $M(t)$ and the numerator is $-M'(t)$ and $K(t) = \ln(M(t))$, we get the following relation,

$$-\bar{r}(t) = \frac{M'(t)}{M(t)} = K'(t). \tag{2}$$

Hence, the first derivative of $-\bar{r}(t)$ is the second cumulant (i.e. the variance) of $-R$, the second derivative of $-\bar{r}(t)$ is the third cumulant of $-R$ (related to the skewness), the third derivative is the fourth cumulant (related to excess kurtosis), and so on. Since $-R$ can be viewed as fitness, the above is essentially equivalent to a generalization of Fisher's fundamental theorem of natural selection, according to which the *n*-th derivative of mean fitness over time is the *n*+1 cumulant (28, 29).

Since in most cases the force of infection is not constant, we further extend this relation between the hazard rates and cumulants. If we let $r_i = FOI(t) \cdot q_i$ where $r_i$ is individual *i*'s hazard rate, *FOI* is the force of infection at time *t*, and $q_i$ is the individual's relative hazard, we can recover the above relation in terms of a transformation of time $s = \int_0^t FOI(\tau)d\tau$,

$$-\bar{r}(s) = \frac{M'_2(s)}{M_2(s)} = K'_2(s) \tag{3}$$

where $K_2$ is the cumulant generating function for the distribution of $-q_i$.

We now consider the hazard ratio comparing vaccinated to unvaccinated individuals, $HR(t) = \bar{r_v}/\bar{r_u}$, where $\bar{r_v}$ is the average hazard rate in not-yet-infected vaccinated individuals and $\bar{r_u}$ the analogous for unvaccinated individuals. To find the rate of change of the hazard ratio and recalling that the first derivative of the mean hazard rate is the second cumulant (i.e. variance) of the hazard rates, we apply the quotient rule (or the quotient and chain rules for the case of time-varying *FOI*) which yields the following equation

$$\frac{d}{dt}[HR] = \frac{-\sigma_v^2 \bar{r_u} + \sigma_u^2 \bar{r_v}}{\bar{r_u}^2}, \tag{4}$$

where $\sigma_v^2$ is the variance of the vaccinated group's hazard rates and $\sigma_u^2$ is the variance of the unvaccinated group's hazard rates.

If at time *t*=0 underlying risk is distributed Gamma($\alpha_\gamma$, $\beta_\gamma$) and vaccine protection Beta($\alpha_\beta$, $\beta_\beta$) and the two are independent, $\alpha_\gamma - \alpha_\beta - \beta_\beta$ determines the direction in which these heterogeneities affect HR(*t*) as

$$\text{sign}[HR'(0)] = \text{sign}(\alpha_\gamma - \alpha_\beta - \beta_\beta). \tag{5}$$

Hence, even in this simple scenario, heterogeneity can cause either an increase or decrease in mVE.

### 3.4 Vaccine Effectiveness Under Competing Heterogeneities

We consider the interplay of heterogeneities in both underlying frailty and vaccine protection and compare vaccine effectiveness estimates to our predicted value based on Equation 4. We find that overall estimates tend to oscillate around the predicted value (purple dashed), as seen in Figure 3. Here we find that, depending on the underlying distributions, the mVE can increase, decrease, or remain steady around $VE_{NE}$, the vaccine efficacy under no epidemic. Likewise, the extent of observed change is dependent on the interplay of both distributions with some changing a negligible amount and others shown here changing >20% compared to the $VE_{NE}$ value. Furthermore, for all of the combinations shown in Figure 3, we maintained a $VE_{NE}$ of 50%, but the initial protection level given also mediates how far from $VE_{NE}$ a given distribution can change, as seen in Figure S2.

Even larger changes of mVE can be found by either extending the study period, allowing for more individuals to get infected and thus contribute to the over- or underestimation, or by considering a more extreme distribution. Additionally, it is theoretically possible to generate non-monotonic behavior, as shown in Figure S3, where mVE can go both up and down from the $VE_{NE}$ value.

### 3.5 Modeling Waning Protection

Without direct challenge studies, estimating heterogeneity in vaccine effectiveness can be fraught. However, many studies use antibody titers as a correlate of protection, including those for SARS-CoV-2 (30-32). Using data on waning SARS-CoV-2 antibodies, we created a distribution for initial protection in a population that then wanes over time. We model waning of antibodies as a power law of the form

$$Ab = C[(t+41)/42]^{-1} \qquad (6)$$

in line with our previous studies (9-11). The exponent -1 corresponds to relatively fast waning. Here C for each individual is drawn from a log normal distribution with a standard deviation of 0.75-1.5 natural log in line with (11, 33, 34). In previous studies, we analyzed antibodies and waning starting at day 42 post-infection or vaccination. We assume here that all individuals in the vaccinated group are fully vaccinated before the study begins. After which we correlate an individual's antibody level to their individual VE using

$$1 - VE = \min[Ab^{-1/2}, 1], \qquad (7)$$

where the antibody to VE conversion exponent is based on an adjustment to the relationships given in (34) for HAI titers and risk of infection with exponents of approximately -0.35. Because HAI titer is only one component of the antibody response, we slightly increased the strength of the relationship and used -0.5. We call this the *risk-correlate model*.

We also consider a different relationship between VE and antibody based on a within-host stochastic extinction model

$$1 - VE = \max\left[\frac{1 - exp\left(\frac{m}{R_0} - \frac{m \cdot a}{a + k \cdot Ab}\right)}{1 - exp\left(\frac{m}{R_0} - m\right)}, 0\right], \qquad (8)$$

where here $a = 10$ is the death rate of virions, $R_0 = 10$ is the basic reproductive number at the between-cell level (35), $k \cdot Ab$ represents the scaled level of antibody, and $m = 0.5$ is the product of the number of viral particles per inoculum and a virion's probability of successfully infecting a cell in the absence of antibodies (see Supplement for derivation and additional details). This relationship gives qualitatively similar results to the risk-correlate model.

Antibody is scaled to give an approximately 90% initial vaccine protection in the population when the standard deviation for natural log of antibody is equal to 1, again in line with (11, 33, 34); inherently, as standard deviation is varied, this causes the initial average of vaccine-induced protection in the population to vary slightly. This distribution replaces the beta distributions used in Section 3.4 for vaccine protection while underlying frailty in both groups continues to be modeled with gamma distributions with CoV based on (36) who estimate CoV of 0.7-1.5 (mean of 0.9) based on contact surveys of very short duration (e.g. two days) and CoV of 0.3-0.9 (mean of 0.5) based on contact surveys when aggregating by 1 year age categories. As elaborated by (36), the former is likely an over estimate whereas the latter is likely an underestimate, hence we consider CoV of 0.5-1.

Using the risk-correlate model, Figure 4 compares a simulation without any heterogeneity (Figure 4A) to simulations with just heterogeneity in (antibody-induced) vaccine protection or underlying frailty and simulations with heterogeneity in both protection and underlying frailty, where underlying frailty is characterized by the coefficient of variation (CoV) of the gamma distribution where $CoV=1/\sqrt{\alpha_\gamma}$ and the mean of the gamma distributions are held the same as the previous figure. Recapitulating the earlier simulations, heterogeneity in vaccine protection results in an increase relative to $VE_{NE}$. However, heterogeneity in underlying frailty overwhelms this positive trend and causes VE estimates to be underestimated. Similar qualitative results are also given using the unadjusted power law exponent estimated from the HAI titers (a likely underestimate) as shown in the Supplement.

Using the within-host stochastic model given by Equation 8 yields similar results, as seen in Figure 5 to the risk-correlate model. For both models the degree of the over- or underestimation (relative to $VE_{NE}$) at the end of the season is given in Table S1. Again, for plausible acute infectious disease parameters, mVE tends to be approximately the same as or underestimates $VE_{NE}$.

## 4 Discussion

Heterogeneity complicates the ability to accurately estimate the extent of vaccine-induced protection in a population as well as if protection is truly waning or merely appears to be waning. While this has been extensively investigated for underlying frailty (4-6) the confounding effect of heterogeneity in vaccine protection has been less thoroughly explored. In many cases (7, 37, 38), only all-or-nothing and leaky vaccines are investigated but we argue that these are edge cases that are wonderful for illustrating theory but are unlikely to accurately model real-world responses to vaccination. In particular, this study considers a much wider array of distributions and shows that the net effect of selection on these heterogeneities can cause either an increase or a decrease in mVE with the effect given by concise and interpretable formulas.

We parameterize our model using data from epidemiological and immunological studies and also incorporate within-host modeling of the immune system and pathogen. We find that, within the estimated ranges, mVE is likely to be underestimated; however, the degree of underestimation is quite varied with heterogeneity in vaccine response offsetting anywhere from <10% to >100% (median of 29%) of the effect of heterogeneity in underlying susceptibility alone (Table S1). Therefore, vaccine effectiveness estimates should be interpreted with caution, especially over time as the heterogeneities continue to accumulate differential outcomes. While mVE seems more likely to underestimate than overestimate VE$_{NE}$, underestimation should not be assumed as our range of plausible parameters includes cases without any underestimation.

Previous studies have used all or nothing or beta distributions to model vaccine induced protection (7, 37, 38). However, modeling waning with such distributions is not straightforward. Our technique of modeling decay of immune responses (in our case, antibodies) at the individual level and converting these immune responses into individual level VE is more transparent and possibly easier to implement than modeling waning by shifting a beta distribution over time.

There are some important caveats to interpreting our results. Although Equations 2-4 give the effect of selection on the hazard rates and hazard ratios, the hazard rates and ratios are also affected by regression towards (or away from) the mean. Regression towards the mean would tend to mitigate the effects of both heterogeneities. Secondly in our simulations heterogeneity in underlying susceptibility and heterogeneity in vaccine response are uncorrelated at baseline. Allowing for correlations permits for more diverse outcomes and affects not only waning but also the initial level of mVE. It should be noted that Equations 2-4 are valid even in the presence of such correlations. Extending our simulations to include such correlations and also regression towards or away from the mean is straightforward. In the current study, we focused only on the hazard ratio as a measure of VE. We do not consider the effect of seasonality, spatial structure, or epidemic waves. We also did not consider all possible combinations of distributions, but rather limited ourselves to those implied by data for acute respiratory infections.

Overall, we give concise statistical formulas for understanding the effect of selection on mVE and give estimates of the magnitude of this effect under a variety of situations based on both antibody and frailty data (10, 36). Our results suggest that VE$_{NE}$ is likely but not certain to be higher than mVE due to variation at the individual level and that the level of discrepancy is dependent on the specifics of the population and vaccine meaning that a simple overall correction cannot be applied. Further exploration for how to correct for these factors statistically or via study design is essential to more accurately understanding vaccine-induced protection.

## 5  Conflict of Interest:

The authors declare that the research was conducted in the absence of any commercial or financial relationships that could be construed as a potential conflict of interest.

## 6  Author Contributions

AN, HA, and VZ contributed to the conception and design of this study. AN performed the simulations. HA designed the within-host models. AN, HA, and VZ analysed the simulation results. AN and HA wrote the first draft of the manuscript. All authors contributed to the manuscript revision, read, and approved the submitted version.

## 7  Funding

This work was supported by the National Heart, Lung, and Blood Institute and the National Institute of Allergy and Infectious Diseases, National Institutes of Health (grants U01 HL139483, U01 AI150747, and U01 AI144616).

## 8  Data Availability Statement

The code used to generate the simulations for this study as well as for its analysis can be found upon publication at the Zarnitsyna Lab Github [https://github.com/ZarnitsynaLab/ArielNikas-VEHeterogeneity].

## 10 Figure Legends

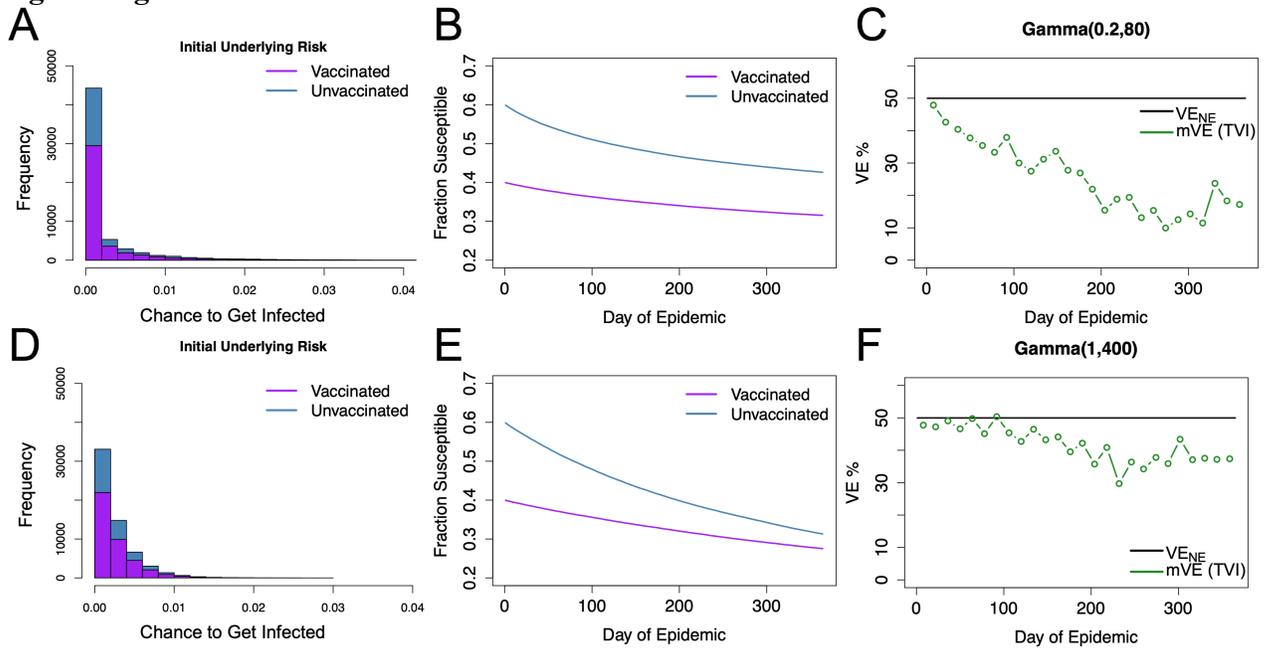

**Figure 1: Gamma distributed underlying risk with vaccine protection (VE) homogeneous at 50% protection.** For underlying risk described by Gamma(0.2, 80), the distribution of the unvaccinated and vaccinated population's daily risk is given in blue and purple as shown in Panel A. Panel B shows the fractions of the susceptible individuals in each group. Panel C shows the estimated vaccine effectiveness (mVE) which drops markedly below the given level we expect of the leaky vaccine (black), decreasing to 17% from the original 50%. Panels D-F, display the same results for Gamma(1, 400). As seen in F, the estimated vaccine effectiveness is below the true value but is not as severe as Panel C, only decreasing to 38.5%.

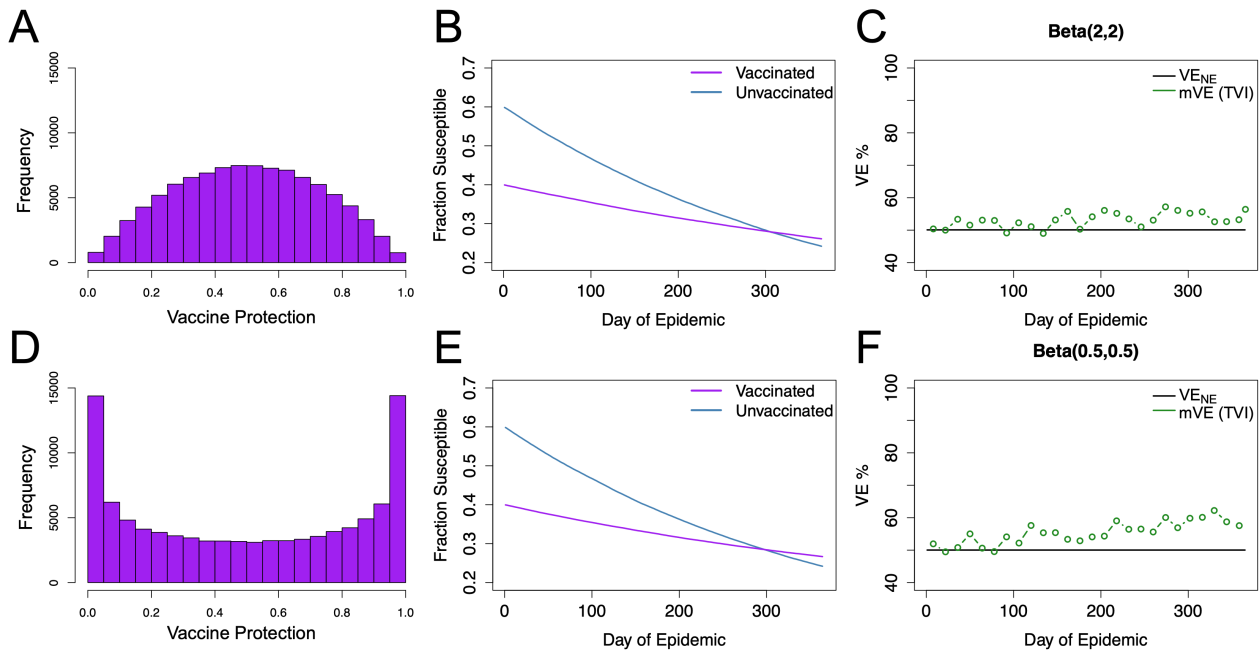

**Figure 2: Beta distributed vaccine protection with homogeneous underlying risk.** Panels A-C, give the results for vaccine protection distributed Beta(2,2), where Panel A displays the resultant

distribution in the vaccinated population, Panel B shows the change in susceptible populations (as a fraction of total population) over time, and Panel C shows the vaccine effectiveness (mVE) estimate which increased 6%. Keeping the mean the same but changing the distribution shape, as seen in Panel D, to Beta(0.5, 0.5), we likewise see similar infection dynamics but higher mVE, increasing by approximately 10%. Here, $VE_{NE}$ is the vaccine efficacy if there was no epidemic and vaccine protection is constant and leaky at 50%.

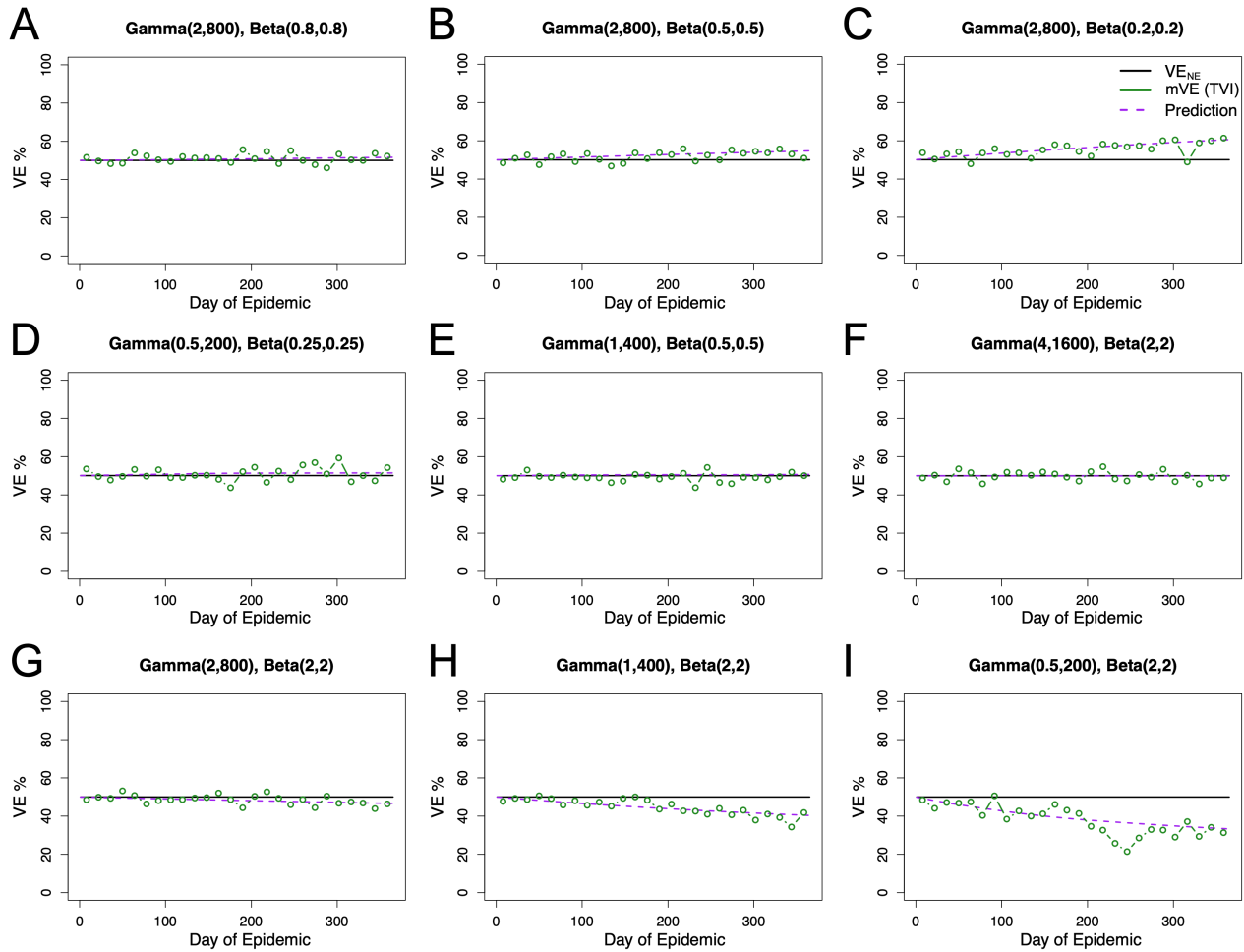

**Figure 3: Competing heterogeneities allow for diverse outcomes in mVE.** Per Equation 5, we predict an increase in observed vaccine effectiveness in Panels A-C, no change in Panels D-F, and a decrease in Panels G-I. For all panels our predicted value (purple dashed) closely matches the mVE (green). Given that an individual's vaccine protection is constant for these simulations (black), this reiterates the difficulty in interpreting changes in mVE.

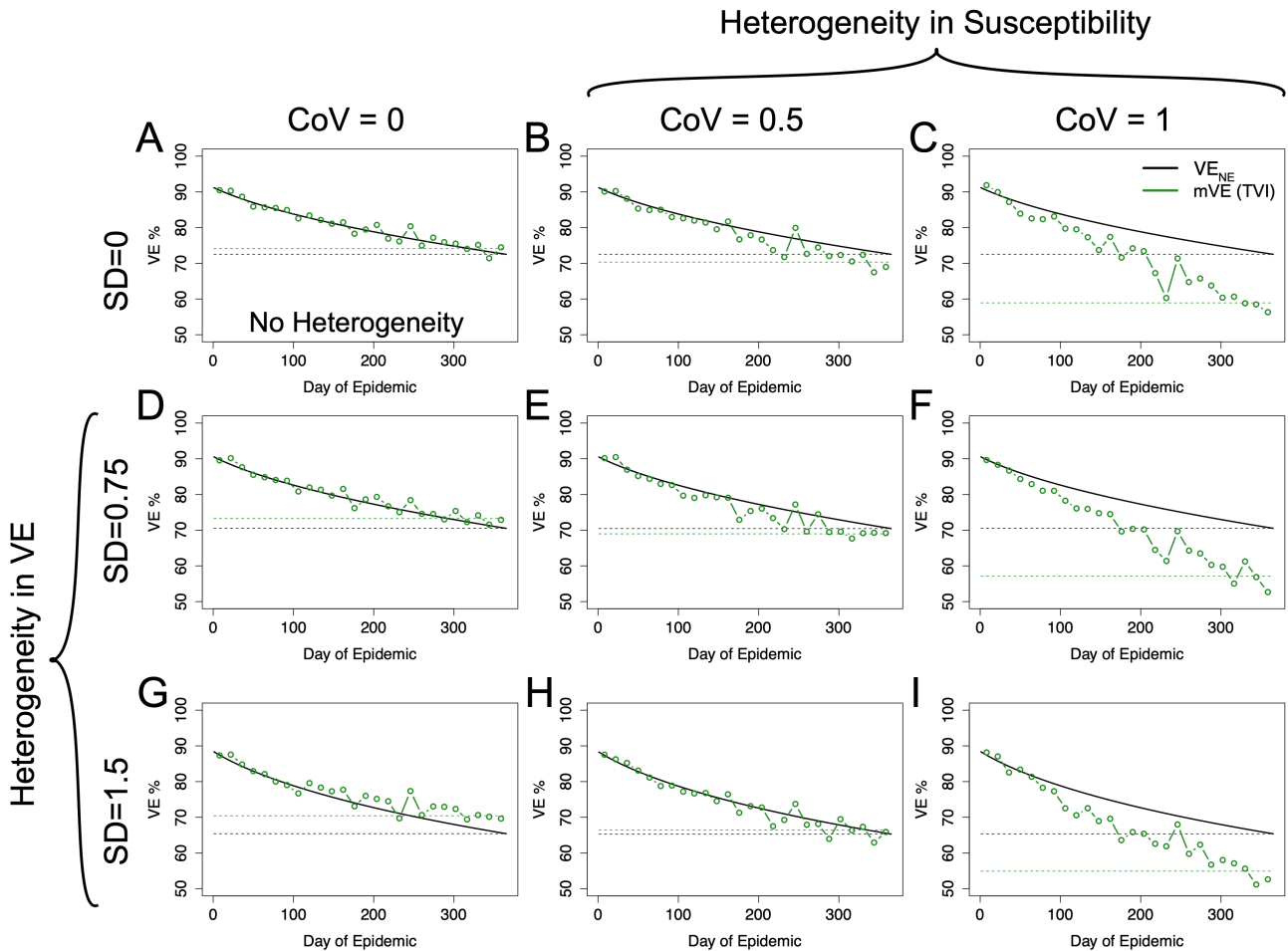

**Figure 4: mVE for plausible acute infectious disease parameters is likely to underestimate VE$_{NE}$ but not necessarily in all circumstances.** Under no heterogeneities, as in Panel A, mVE is extremely close to VE$_{NE}$; however, the introduction of heterogeneity biases the estimate. In the first column, showing simulations lacking heterogeneity in underlying susceptibility, heterogeneity in antibody biases the estimate upwards. In the first row, without any heterogeneity in antibody, the bias is downwards. With both, the underlying heterogeneity in susceptibility outcompetes heterogeneity in VE and leads to an underestimate relative to VE$_{NE}$, though not as extreme as would be seen if a vaccine was purely homogeneously leaky except in Panel H where the two effects approximately cancel each other out. SD indicates the standard deviation of antibody (at a given time) in natural logs; higher SD in antibody translates to higher variability in vaccine protection via Equation 6. CoV indicates the coefficient of variation in underlying susceptibility at the beginning of the simulation.

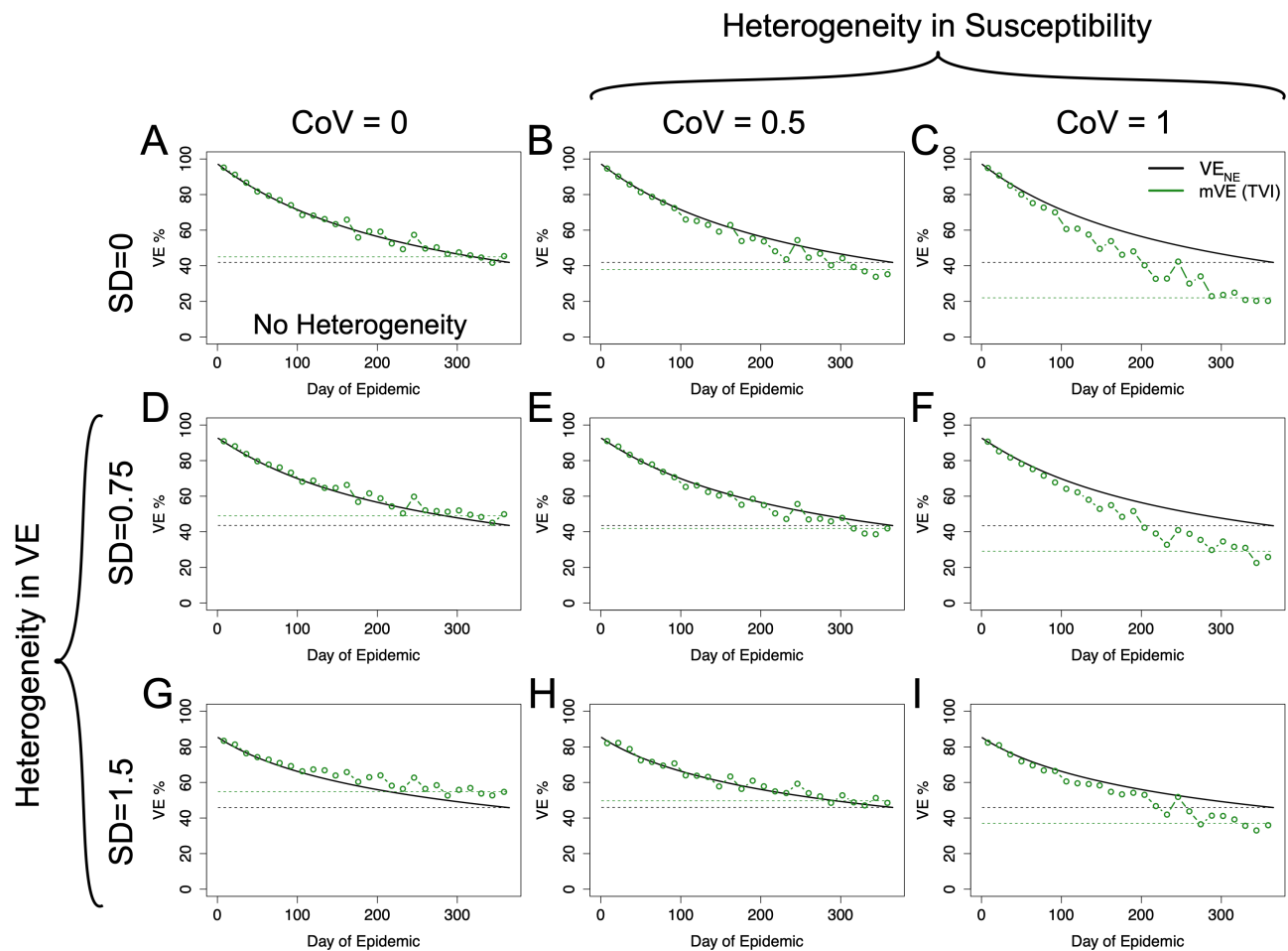

**Figure 5: Within-host stochastic model.** Utilizing the within-host stochastic model (Equation 8) for antibody-mediated, vaccine-induced protection yields similar results to the risk-correlate model (Equation 7). As expected, increasing heterogeneity in underlying susceptibility (moving left to right) pushes mVE downwards while increasing heterogeneity in vaccine-induced protection (moving top to bottom) pushes mVE upwards. When these effects are mixed, as in Panels E, F, H, and I, the heterogeneities compete.

# *Supplementary Material*

**Alternative, Non-symmetric Beta Distributions**

While the beta distributions considered in the main text were selected to reflect potential heterogeneity under vaccination, such that the vaccine either protected most people reasonably similarly (Beta(2, 2)) or gave a wide range of protection with some people responding strongly and others not (Beta(0.5, 0.5)), we also tested other distributions but they did not qualitatively alter our results. In Figure S1, we compare non-extreme, skewed beta distributions which, while they did both trend upwards, did not have large biases. However, changing these two skewed distributions to be more extreme (Figure S1 G, J), we see greater bias upwards as expected.

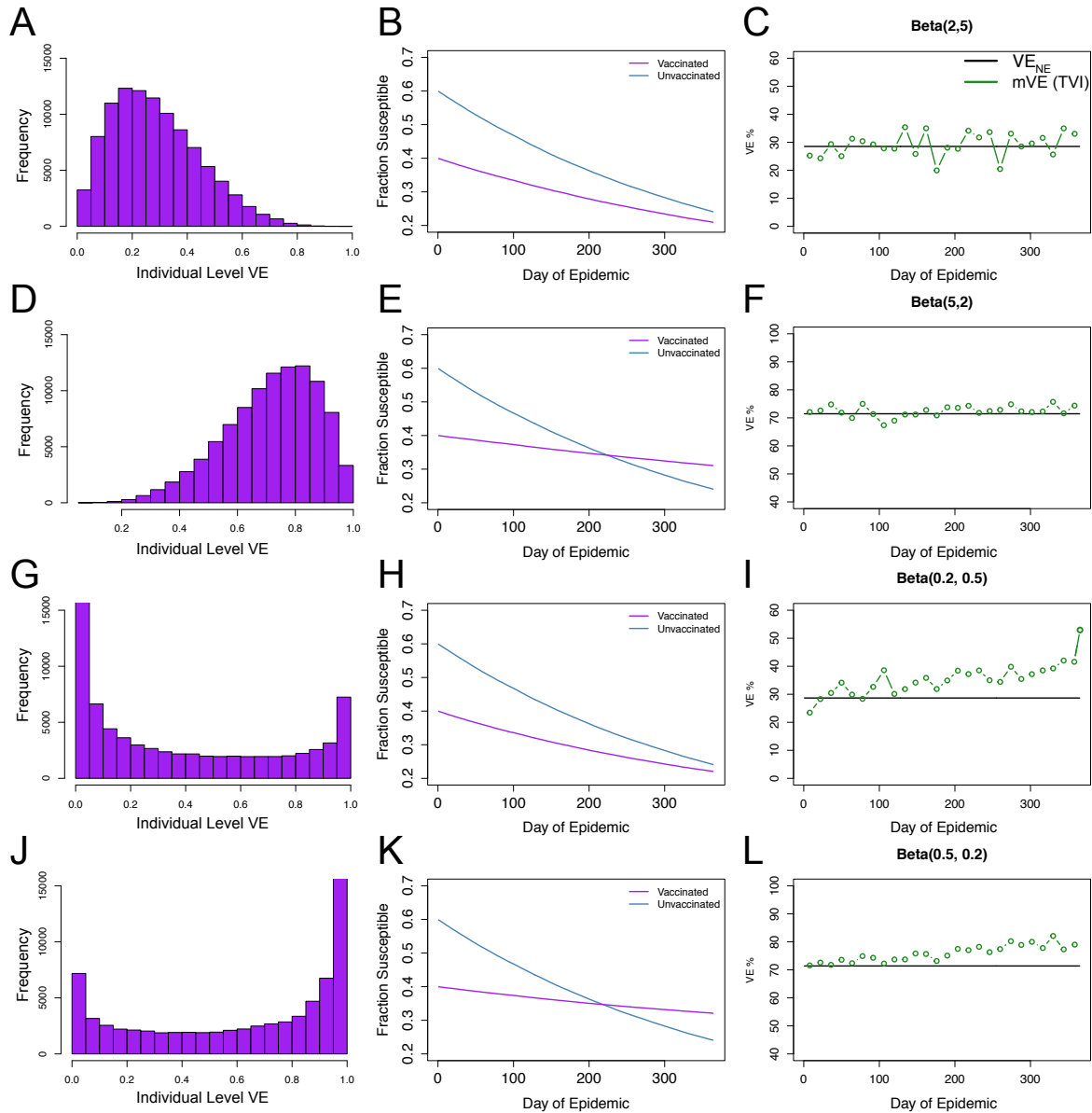

**Figure S1: Non-symmetric beta distributions also bias upwards.** Our results hold for other beta distribution shapes as well. In the first two rows we compare skewed but non-extreme distributions where in the first row the majority of the vaccinated population have little protection and in the second row the majority have high protection. We compare this to a more extreme version of each in the third and fourth rows. While the less extreme distributions remain near though slightly above $VE_{NE}$, the others increase more obviously with one increasing approximately 20%.

## 11 Starting Vaccine-induced Protection Influences the Total Amount of Perceived Waning

The initial level of protection from vaccination, VE(0), influences the amount of perceived waning observed. Extreme VE(0) show less observed change while middle values consistently showed the most. This pattern holds for both heterogeneity in vaccine response and heterogeneity in underlying frailty, as shown in the comparisons in Figures S2. Under no heterogeneity in susceptibility, leaky vaccination caused no more than 1% observed change in mVE in either direction. When $\alpha_\gamma$ is 1, change in mVE ranged from -7 to -17% while when $\alpha_\gamma$ is 2 the change in mVE ranged from -4 to -10% with the largest decreases being at the mid-range values for both. These changes are plotted explicitly in Figure S2.

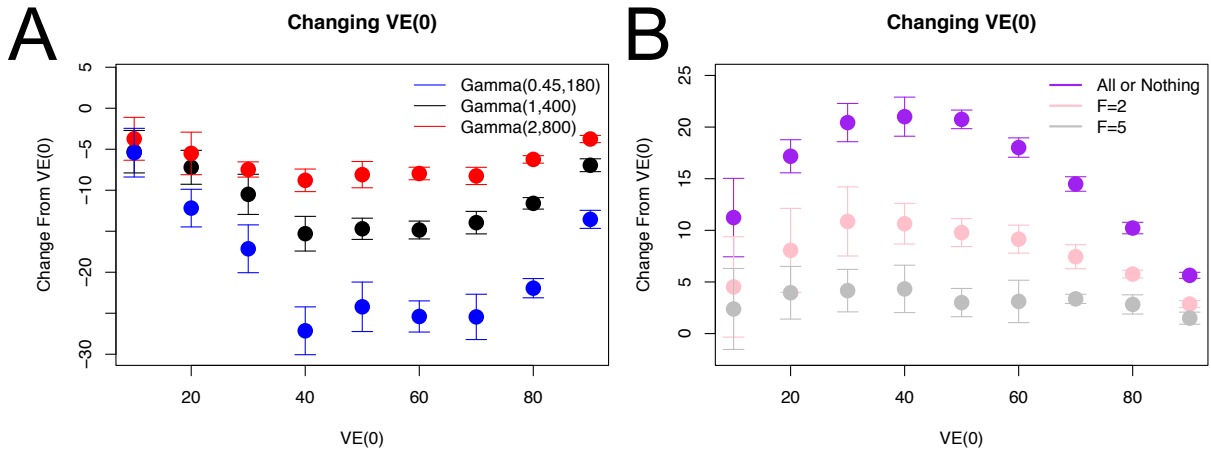

**Figure S2: Effect of starting VE protection level.** Different VE(0) values affect how much mVE can change. In Panel A, when there is no heterogeneity in individual level VE, regardless of chosen gamma distribution mid-range values of VE have the greatest capacity to change based on the difference from the initial value and the mean of the last five time points. In Panel B, if there is no heterogeneity in susceptibility but only in vaccine-induced protection as given by either a beta distribution or all-or-nothing protection, again midrange values have the largest capacity for change but this time bias upwards. For the beta distributions, $F=1+\alpha_\beta+\beta_\beta$ indicates the fold reduction in variance relative to all-or-nothing protection with the same mean. For VE = 50%, the F=2 point corresponds to the main text Figure 2F and the F=5 point corresponds to Figure 2C.

## 12 Predictions Not Necessarily Monotonic

While in the main text a variety of distributions were considered, we also tested additional distribution parameters for both the underlying heterogeneity in susceptibility and vaccine protection. Most of these combinations yielded monotonic predictions for change in VE; however, some, like the one pictured in Figure S3, can be non-monotonic. Here the predicted value decreases after a short period as can be seen by the dashed purple line appearing below the original value and then later increases past the original line around day 225 before dipping down and up yet again.

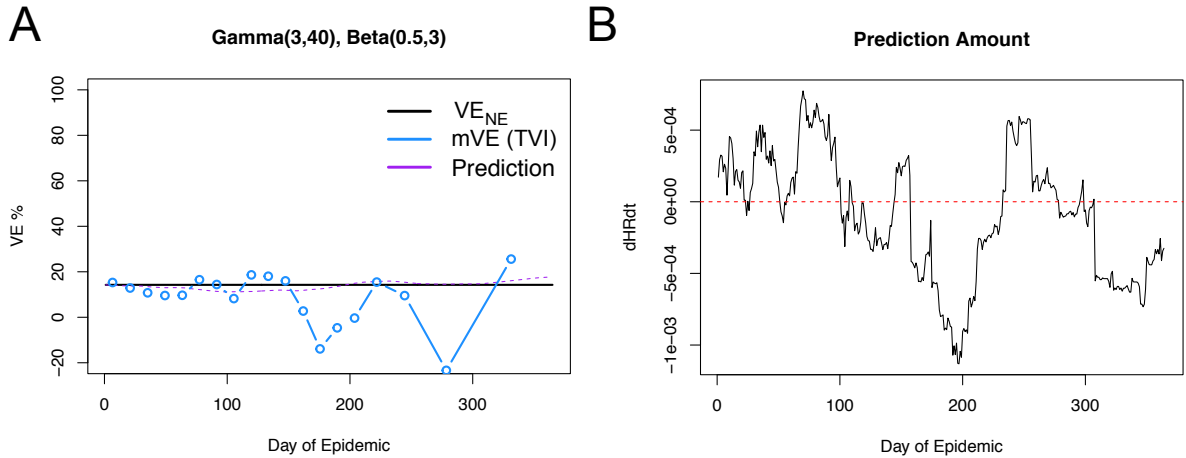

**Figure S3: The observed and predicted vaccine effectiveness are not necessarily monotonic.** While most parameter regimes considered yielded monotonic predictions for VE, that is not necessarily the case for any given parameter combination. Here we show an example where the predicted value both increases and decreases over time. This is especially evident in Panel B, as anytime the predicted change (black) crosses zero (red dashed), the direction of the change switches.

## 13 Unadjusted Antibody Exponent

In the main text, we explain our reasoning for increasing the antibody to VE conversion as HAI-specific titers are likely a lower bound for total antibody. We show results using this lower bound derived from [1] in Figure S4.

Here, instead of the main text conversion equation (Equation 7), we use

$$1 - VE_{NE} = \min[Ab^{-0.35}, 1]. \tag{1}$$

This exponent was approximated from the HAI titers found in [1]. This change did not qualitatively alter results; heterogeneity in vaccine protection continues to bias the VE estimate upwards while heterogeneity in the underlying susceptibility continues to bias the VE estimate downwards and in most scenarios will out-compete vaccine protection's bias.

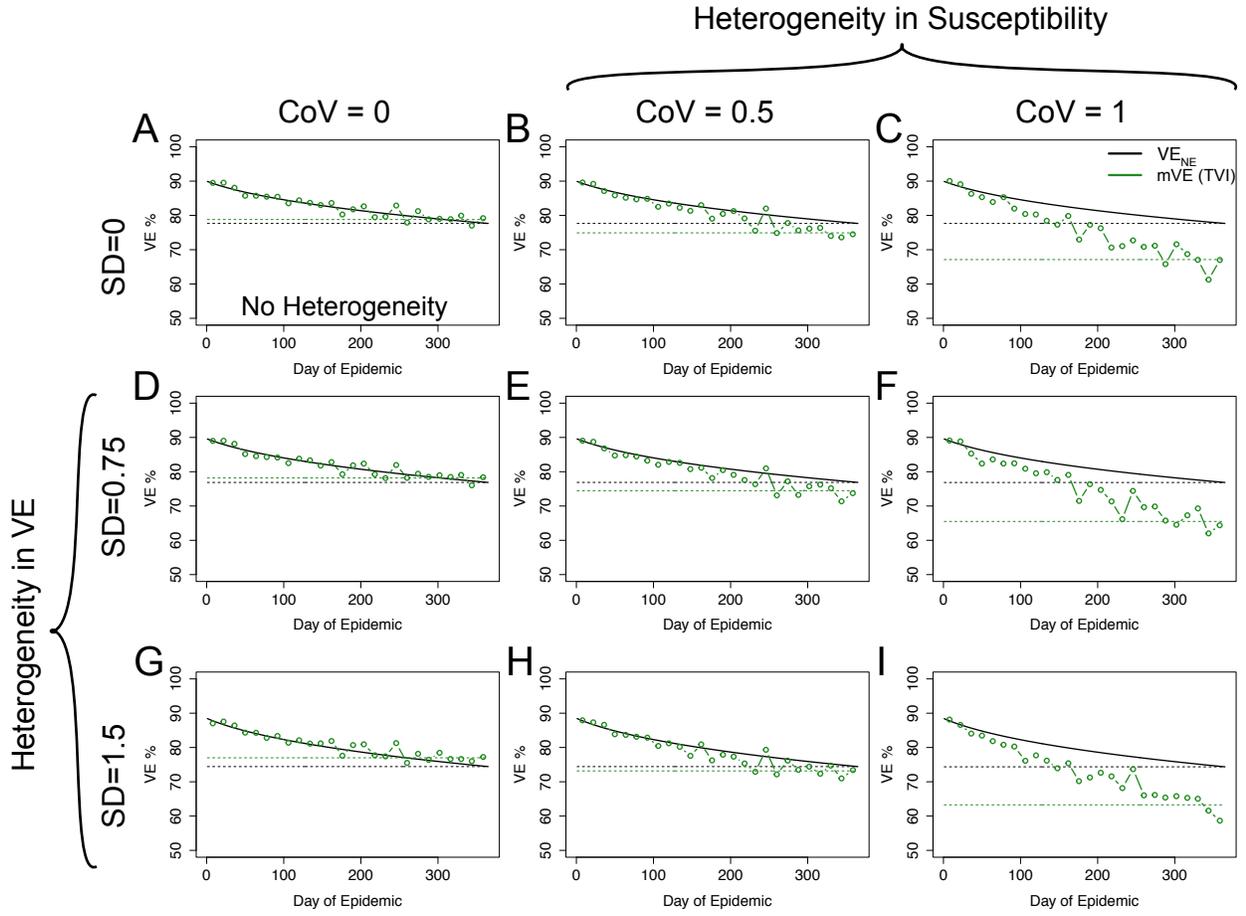

**Figure S4: Unadjusted data estimate: overestimation of waning when frailty is added.** Using the unadjusted power law exponent of -0.35 as estimated from [1] has similar behavior to the adjusted version seen in the main text. Again, increasing heterogeneity in underlying susceptibility (moving left to right) biases mVE downwards while increasing heterogeneity in vaccine-induced protection (moving top to bottom) biases mVE upwards. When these effects are mixed, as in Panels E, F, H, and I, the heterogeneities compete.

## 14 Within-Host Stochastic Model: Derivation

Here, we keep the same level of waning antibodies as given in Equation 6 of the main text, but we additionally consider the chance that an inoculum actually succeeds in causing infection.

To derive, first consider the probability that a single virion infects a cell,

$$P_{inf} = \frac{r}{r+a+k\text{Ab}} \approx \frac{r}{a+k\text{Ab}}, \qquad (2)$$

where $r$ is the rate of infection, $a$ is the rate of viral death, $k \cdot \text{Ab}$ is the rate of clearance by antibody, and $r \ll (a + k \cdot \text{Ab})$. Assuming that a virion successfully infects a cell, we take the probability of early stochastic extinction to be

$$P_{SE} = min[1/R_0^*, 1], \qquad (3)$$

as follows from the assumption that $R_s$, the actual number of cells infected by a given cell early in infection, is Gamma Poisson distributed with $R_s \sim$ Poisson(mean=M) and M~Gamma($\alpha_\gamma$=1, $\beta_\gamma$=$R_0^{*-1}$). Additionally, we assume that $R_0^*$ is proportional to the infectivity of the virus, so

$$R_0^* \approx R_0 \frac{a}{a+k\text{Ab}}, \tag{4}$$

where $R_0$ is the basic reproduction number of a cell in the absence of antibody. Building upon Equations S2 and S3, we then consider an inoculum with $n$ virions where the probability of infection which escapes early stochastic extinction is

$$P_n = max\left[1 - \left(1 - \frac{r}{a+k\text{Ab}}\left(1 - \frac{a+k\text{Ab}}{aR_0}\right)\right)^n, 0\right]$$

$$\approx max\left[1 - exp\left(\frac{r}{a} \cdot n \cdot \left(\frac{-a}{a+k\text{Ab}} + \frac{1}{R_0}\right)\right), 0\right]. \tag{5}$$

This then becomes a hazard ratio $HR_W$ by normalizing the above, yielding

$$HR_W = max\left[\frac{1-exp\left[\frac{m}{R_0} - \frac{m \cdot a}{a+k \cdot \text{Ab}}\right]}{1-exp\left[\frac{m}{R_0} - m\right]}, 0\right], \tag{6}$$

where we set $m = \frac{rn}{a}$.

We take the viral death rate to be $a$=10 in line with [2]. For the basic reproduction number in the absence of antibody, we use $R_0 = 10$ which is in the range estimated in [2].

For our combined parameter $m = \frac{rn}{a}$, we have additional considerations.

In the absence of antibody, the probability of infection taking hold is the denominator of Equation S7. Here, $m$ in (0,1] corresponds to $P_n$ of up to 59% per exposure in naive individuals and hence covers both small and moderately large exposures. As seen in Figure S5 changing $m$ from near 0 to 1 does not result in large changes in the hazard ratio and therefore we chose $m$=0.5 as a representative for this range. If $m$ is extremely large, this causes the hazard ratio to essentially become all-or-nothing.

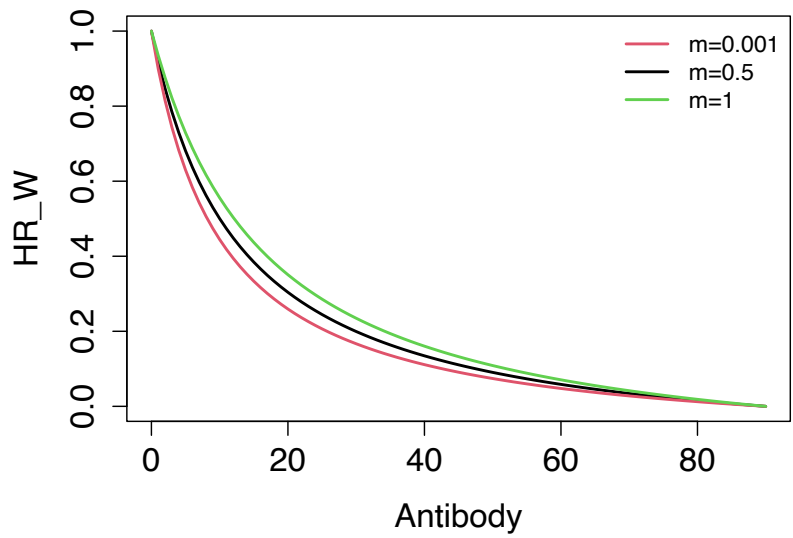

**Figure S5: Effect of *m* on the hazard ratio.** Hazard ratio results are not particularly sensitive to choice of *m* in (0,1] and thus a midrange value was selected (*m*=0.5).

## 15 Comparison of Risk-Correlate and Within-Host Stochastic Model Outcomes

The difference in end-season estimates from $VE_{NE}$ for both the risk-correlate model and the within-host stochastic model are given in Table S1, below.

**Table S1: Average of the difference between mVE and $VE_{NE}$ over the last three time points.** These values indicate the effect size of heterogeneity on mVE in our simulations. Heterogeneity in underlying susceptibility alone led to mVE underestimating $VE_{NE}$ by 3.6%, 8%, 15.4%, and 22.9%. Adding heterogeneity in vaccine response offset anywhere from a negligible (<10%) to >100% (median: 29%) of the effect of heterogeneity in underlying susceptibility alone.

| CoV | SD | Risk-Correlate | Within Host |
|---|---|---|---|
| 0 | 0 | 0.45 | 0.6 |
| 0 | 0.75 | 1.6 | 2.9 |
| 0 | 1.5 | 3.9 | 6.9 |
| 0.5 | 0 | -3.6 | -8.0 |
| 0.5 | 0.75 | -3.6 | -5.0 |
| 0.5 | 1.5 | 2.0 | 2.0 |
| 1 | 0 | -15.4 | -22.9 |
| 1 | 0.75 | -14.4 | -18.2 |
| 1 | 1.5 | -13.1 | -12.0 |